# Analysis of Transition Path Ensemble in the Exactly Solvable Models via Overdamped Langevin Equation

De-Zhang Li[1], Jia-Rui Zeng[1], Wei-Jie Huang[1], Yao Yao[1] and Xiao-Bao Yang[1*]

[1] Department of Physics, South China University of Technology, Guangzhou 510640, China.

[*] Corresponding authors. Correspondence and requests for materials should be addressed to X.-B. Y. (email: scxbyang@scut.edu.cn).

## Abstract

Transition of a system between two states is an important but difficult problem in natural science. In this article we study the transition problem in the framework of transition path ensemble. Using the overdamped Langevin method, we introduce the path integral formulation of the transition probability and obtain the equation for the minimum action path in the transition path space. For the effective sampling in the transition path ensemble, we derive a conditional overdamped Langevin equation. In two exactly solvable models, the free particle system and the harmonic system, we present the expression of the conditional probability density and the explicit solutions for the conditional Langevin equation and the minimum action path. The analytic results demonstrate the consistence of the conditional Langevin equation with the desired probability distribution in the transition. It is confirmed that the conditional Langevin equation is an effective tool to sample the transition path ensemble, and the minimum action principle actually leads to the most probable path.

# 1. Introduction

Transition of a system from an initial state to a final state is one of the most important phenomena in nature. Understanding the pathways and mechanisms of transition phenomena like structural transformation, phase transition and chemical reaction has been of great interest in natural science. Effective approach to the transition problem, especially to the dynamics of rare events, still remains a difficult task.

A famous theoretical framework to study the transition problem is offered by the transition state theory [1-5]. Transition state, usually determined by a saddle point of the potential energy surface, is the key idea of this framework. The transition path generated by the transition state theory is usually the minimum energy path crossing the transition state [6, 7]. Two classes of methods of searching the transition state and minimum energy path on the potential energy surface are widely used, the nudged elastic band method [8, 9] and the energy surface walking methods [10, 11]. However, the minimum energy path does not always work when the transition state is not exactly a saddle point on the potential energy surface [7]. Another popular approach to the most probable path is the minimum action path [12-16]. In this framework, an action as a functional of the transition path should be defined, e.g., the famous Onsager-Machlup action [17, 18]. This action functional is directly related to the statistical weight for a certain transition path. The minimum action path, which has the largest weight associated with the action functional, can then be derived from the variational principle. For the system with complex energy landscape such as material system [19-22], determination of the most probable path is still challenging.

Transition path sampling method [23-29] provides an alternative view to the transition problem. Not limited to a certain path, the dynamic information in the transition is generated from sampling the ensemble in the space of all possible transition paths. The distribution or the statistical weight for a path, like that associated with the action functional mentioned above, plays a key role in the sampling. Techniques such as Monte Carlo are employed. Since the Langevin approach to the transition problem had been introduced for a long time [30], one will expect the application of Langevin dynamics in transition path sampling. Recently, a overdamped Langevin based method in the framework of transition path ensemble is proposed [31-35], which avoids the drawback of Monte Carlo that generating highly correlated trajectories. This method adopts the concept of Langevin bridge in the mathematical literatures [36, 37]. The transition probability of the system driven by overdamped Langevin equation is related to the propagator in the path integral formulation [38-44] naturally. Given the initial state, the final state and the transition time, a conditional overdamped Langevin equation is constructed as a stochastic description of the transition paths. Hence, it provides an effective stochastic differential equation to sample the transition path ensemble.

In this article, we follow closely the derivations in [31-35] to analyse the transition path ensemble and most probable path of two exactly solvable models, the free particle system and the harmonic system. For the sake of simplicity, we discuss the result of one-dimensional system. The extension to general multi-dimensional case is straightforward. In Section 2, we introduce the path integral formulation of the transition probability and show the equation for the minimum action path. A conditional overdamped Langevin equation, which produces the desired probability distribution in the transition, is obtained in Section 3. The detailed analyses of the effective probability distribution, Langevin equation and the most probable path for two exactly solvable models are presented in Section 4. Conclusions are outlined in Section 5.

## 2. Path integral formulation

Overdamped Langevin equation is suitable for transition problem, compared to the time scale of the usual underdamped Langevin equation. In this article we study the system driven by overdamped Langevin equation in the form

$$\frac{dx}{dt} = -\frac{1}{m\gamma}\frac{\partial U}{\partial x} + \sqrt{\frac{2}{\beta m\gamma}}\eta(t) \ . \tag{1}$$

Here $m$ is the mass, $U(x)$ is the potential energy, $\gamma$ is the friction coefficient and $\beta = 1/k_B T$ with the Boltzmann constant $k_B$ and the temperature $T$. $\eta(t)$ is the white noise random variable associated with the Wiener process satisfying

$$\langle \eta(t) \rangle = 0, \ \langle \eta(t)\eta(t') \rangle = \delta(t-t') \ . \tag{2}$$

The evolution of the system is a Markovian process. The term $\sqrt{\frac{2}{\beta m\gamma}}$ associated with $\eta(t)$ in Eq. (1) fulfils the fluctuation-dissipation relation, which guarantees that the overdamped Langevin equation generates the Boltzmann distribution in the stationary state. Rather than the stationary distribution, our interest is the nonstationary process and the corresponding time-dependent probability density $\rho(x,t)$. $\rho(x,t)$ evolves according to the well-known Fokker-Planck equation [45-50]

$$\frac{\partial}{\partial t}\rho(x,t) = \frac{\partial}{\partial x}\left[\frac{1}{m\gamma}\frac{\partial U}{\partial x}\rho(x,t) + \frac{1}{\beta m\gamma}\frac{\partial \rho(x,t)}{\partial x}\right] \ . \tag{3}$$

Because the Fokker-Planck equation is closely related to the Schrödinger equation, we follow van Kampen [51] to define a wave function

$$\psi(x,t) = e^{\beta U(x)/2}\rho(x,t) \ . \tag{4}$$

Then the Fokker-Planck equation Eq. (3) leads to

$$\frac{\partial}{\partial t}\psi(x,t) = \left\{\frac{1}{\beta\gamma m}\frac{\partial^2}{\partial x^2} - \frac{\beta}{4\gamma m}\left[\left(\frac{\partial U}{\partial x}\right)^2 - \frac{2}{\beta}\frac{\partial^2 U}{\partial x^2}\right]\right\}\psi(x,t) \ . \tag{5}$$

This transformation had been studied for several models [52, 53]. It is straightforward to see that Eq. (5) is an imaginary time Schrödinger equation with the Hamiltonian

$$\hat{H} = -\frac{1}{\beta\gamma m}\frac{\partial^2}{\partial x^2} + \frac{\beta}{4\gamma m}\left[\left(\frac{\partial U}{\partial x}\right)^2 - \frac{2}{\beta}\frac{\partial^2 U}{\partial x^2}\right] \ . \tag{6}$$

The corresponding kinetic energy and potential energy operators are

$$\hat{T} = -\frac{1}{\beta\gamma m}\frac{\partial^2}{\partial x^2}, \quad \hat{V} = \frac{\beta}{4\gamma m}\left[\left(\frac{\partial U}{\partial x}\right)^2 - \frac{2}{\beta}\frac{\partial^2 U}{\partial x^2}\right]. \tag{7}$$

The propagator from the time evolution equation $\frac{\partial}{\partial t}\psi(x,t) = -\hat{H}\psi(x,t)$ can then be expressed as $\langle x_e | e^{-t_{tr}\hat{H}} | x_0 \rangle$ with the initial state $x_0$, final state $x_e$ and the transition time $t_{tr}$. To illustrate the relation between the transition probability $\rho(x_e, t_{tr} | x_0, 0)$ and the propagator $\langle x_e | e^{-t_{tr}\hat{H}} | x_0 \rangle$, one should only compare the following two statements

$$\begin{aligned}\psi(x_e, t_{tr}) &= \int \langle x_e | e^{-t_{tr}\hat{H}} | x_0 \rangle \psi(x_0, 0) dx_0 \\ \rho(x_e, t_{tr}) &= \int \rho(x_e, t_{tr} | x_0, 0) \rho(x_0, 0) dx_0\end{aligned} \tag{8}$$

and use the definition of $\psi$ in Eq. (4). For a Markovian process, $\rho(x_e, t_{tr} | x_0, 0)$ is not a functional of $\rho(x_0, 0)$, therefore

$$\rho(x_e, t_{tr} | x_0, 0) = e^{\beta[U(x_0) - U(x_e)]/2} \langle x_e | e^{-t_{tr}\hat{H}} | x_0 \rangle. \tag{9}$$

The path integral formulation for the transition probability $\rho(x_e, t_{tr} | x_0, 0)$ can be conveniently obtained from Eq. (9). The imaginary time propagator can be expressed in the famous Lagrangian form

$$\langle x_e | e^{-t_{tr}\hat{H}} | x_0 \rangle = \int_{x_0}^{x_e} \mathcal{D}[x(s)] \exp\left\{-\frac{\beta\gamma}{2}\int_0^{t_{tr}}\left[\frac{1}{2}m\left(\frac{dx}{ds}\right)^2 + \frac{1}{2\gamma^2 m}\left(\left(\frac{\partial U}{\partial x}\right)^2 - \frac{2}{\beta}\frac{\partial^2 U}{\partial x^2}\right)\right]ds\right\}. \tag{10}$$

The resulting path integral formulation for $\rho(x_e, t_{tr} | x_0, 0)$ is

$$\rho(x_e, t_{tr} | x_0, 0) = e^{\beta[U(x_0) - U(x_e)]/2} \int_{x_0}^{x_e} \mathcal{D}[x(s)] \exp\{-S[x(s)]\} \tag{11}$$

with the action $S$ for a path starting at $x_0$ and ending at $x_e$ at time $t_{tr}$ defined by

$$S[x(s)] = \frac{\beta\gamma}{2}\int_0^{t_{tr}} L\, ds,\tag{12}$$

$$L = \frac{1}{2}m\left(\frac{dx}{ds}\right)^2 + \frac{1}{2\gamma^2 m}\left(\left(\frac{\partial U}{\partial x}\right)^2 - \frac{2}{\beta}\frac{\partial^2 U}{\partial x^2}\right).\tag{13}$$

Here $L$ is the Lagrangian. This formulation agrees with the Itô path integral representation of Langevin equation [34]. Now we see the statistical weight for a path $x(s)$ in the path space is $P[x(s)] \propto e^{-S[x(s)]}$, which can be seen as the distribution of the transition path ensemble. The most probable path in this formulation is thus the path with the minimal action $S$. It is easy to determine the equation for minimum action path from the variational principle, very like that for the Euler-Lagrange equation in classical mechanics. It takes the form

$$\frac{d^2 x(t)}{dt^2} = \frac{1}{\gamma^2 m^2}\left(\frac{\partial U}{\partial x}\cdot\frac{\partial^2 U}{\partial x^2} - \frac{1}{\beta}\frac{\partial^3 U}{\partial x^3}\right)\tag{14}$$

under the boundary conditions

$$x(0) = x_0,\; x(t_{tr}) = x_e.\tag{15}$$

The minimum action path in this formulation is also called the dominant path in some literatures [14, 15, 34].

## 3. Conditional overdamped Langevin equation

Sampling the transition path ensemble with the distribution $P[x(s)]$ via Langevin equation requires an effective modification which includes the constraint $x(t_{tr}) = x_e$. We follow the derivations in [31-35], to construct a conditional overdamped Langevin equation for the desired probability distribution.

Under the condition of starting at $x_0$ and ending at $x_e$ at time $t_{tr}$, the probability density of the system at $x$ at time $t$ $(0 \le t \le t_{tr})$ is

$$\mathcal{P}(x,t) = \frac{\rho(x_\mathrm{e},t_\mathrm{tr}|x,t)\rho(x,t|x_0,0)}{\rho(x_\mathrm{e},t_\mathrm{tr}|x_0,0)} \ . \tag{16}$$

Consider Eq. (9), we may express the conditional probability density using the propagator as

$$\mathcal{P}(x,t) = \frac{\langle x_\mathrm{e}|e^{-(t_\mathrm{tr}-t)\hat{H}}|x\rangle\langle x|e^{-t\hat{H}}|x_0\rangle}{\langle x_\mathrm{e}|e^{-t_\mathrm{tr}\hat{H}}|x_0\rangle} \ . \tag{17}$$

The boundary conditions can be easily verified

$$\mathcal{P}(x,0) = \delta(x-x_0), \quad \mathcal{P}(x,t_\mathrm{tr}) = \delta(x-x_\mathrm{e}). \tag{18}$$

It is well-known in quantum mechanics that the propagator is a solution of time-dependent Schrödinger equation. That is,

$$\frac{\partial}{\partial t}\langle x|e^{-t\hat{H}}|x_0\rangle = \langle x|e^{-t\hat{H}}(-\hat{H})|x_0\rangle = \left\{\frac{1}{\beta\gamma m}\frac{\partial^2}{\partial x^2} - \frac{\beta}{4\gamma m}\left[\left(\frac{\partial U}{\partial x}\right)^2 - \frac{2}{\beta}\frac{\partial^2 U}{\partial x^2}\right]\right\}\langle x|e^{-t\hat{H}}|x_0\rangle, \tag{19}$$

$$\frac{\partial}{\partial t}\langle x_\mathrm{e}|e^{-(t_\mathrm{tr}-t)\hat{H}}|x\rangle = \langle x_\mathrm{e}|e^{-(t_\mathrm{tr}-t)\hat{H}}\hat{H}|x\rangle = \left\{-\frac{1}{\beta\gamma m}\frac{\partial^2}{\partial x^2} + \frac{\beta}{4\gamma m}\left[\left(\frac{\partial U}{\partial x}\right)^2 - \frac{2}{\beta}\frac{\partial^2 U}{\partial x^2}\right]\right\}\langle x_\mathrm{e}|e^{-(t_\mathrm{tr}-t)\hat{H}}|x\rangle. \tag{20}$$

Then the time evolution equation of $\mathcal{P}(x,t)$ can be easily obtained

$$\begin{aligned}\frac{\partial}{\partial t}\mathcal{P}(x,t) &= \frac{\left[\frac{\partial}{\partial t}\langle x_\mathrm{e}|e^{-(t_\mathrm{tr}-t)\hat{H}}|x\rangle\right]\langle x|e^{-t\hat{H}}|x_0\rangle + \langle x_\mathrm{e}|e^{-(t_\mathrm{tr}-t)\hat{H}}|x\rangle\left[\frac{\partial}{\partial t}\langle x|e^{-t\hat{H}}|x_0\rangle\right]}{\langle x_\mathrm{e}|e^{-t_\mathrm{tr}\hat{H}}|x_0\rangle} \\ &= \frac{1}{\beta\gamma m}\frac{\left[-\frac{\partial^2}{\partial x^2}\langle x_\mathrm{e}|e^{-(t_\mathrm{tr}-t)\hat{H}}|x\rangle\right]\langle x|e^{-t\hat{H}}|x_0\rangle + \langle x_\mathrm{e}|e^{-(t_\mathrm{tr}-t)\hat{H}}|x\rangle\left[\frac{\partial^2}{\partial x^2}\langle x|e^{-t\hat{H}}|x_0\rangle\right]}{\langle x_\mathrm{e}|e^{-t_\mathrm{tr}\hat{H}}|x_0\rangle} \\ &= \frac{1}{\beta\gamma m}\frac{\partial}{\partial x}\frac{\left[-\frac{\partial}{\partial x}\langle x_\mathrm{e}|e^{-(t_\mathrm{tr}-t)\hat{H}}|x\rangle\right]\langle x|e^{-t\hat{H}}|x_0\rangle + \langle x_\mathrm{e}|e^{-(t_\mathrm{tr}-t)\hat{H}}|x\rangle\left[\frac{\partial}{\partial x}\langle x|e^{-t\hat{H}}|x_0\rangle\right]}{\langle x_\mathrm{e}|e^{-t_\mathrm{tr}\hat{H}}|x_0\rangle} \\ &= \frac{1}{\beta\gamma m}\frac{\partial}{\partial x}\left\{-2\frac{\partial}{\partial x}\left[\ln\langle x_\mathrm{e}|e^{-(t_\mathrm{tr}-t)\hat{H}}|x\rangle\right]\times\mathcal{P}(x,t) + \frac{\partial\mathcal{P}(x,t)}{\partial x}\right\}.\end{aligned} \tag{21}$$

Naturally, one may compare this equation with that of the unconditional probability Eq. (3) and find that the modified force should be defined as

$$-\frac{\partial \tilde{U}}{\partial x} = \frac{2}{\beta} \frac{\partial}{\partial x} \left[ \ln \left\langle x_e \left| e^{-(t_{tr}-t)\hat{H}} \right| x \right\rangle \right] . \tag{22}$$

Note that the modified force is time-dependent. The resulting conditional Langevin equation is then

$$\frac{dx}{dt} = \frac{2}{\beta m \gamma} \frac{\partial}{\partial x} \left[ \ln \left\langle x_e \left| e^{-(t_{tr}-t)\hat{H}} \right| x \right\rangle \right] + \sqrt{\frac{2}{\beta m \gamma}} \eta(t) . \tag{23}$$

Comparing to the unconditional case Eq. (1) demonstrates that, the modified force guarantees the trajectories end at $x_e$ at time $t_{tr}$. In principle, the probability density of the system driven by this modified Langevin equation is the desired $\mathcal{P}(x,t)$. The trajectories generated from this equation follow the distribution $P[x(s)]$ of transition path ensemble. We will test this conditional overdamped Langevin equation in the free particle system and the harmonic system, of which the modified force in Eq. (22) can be expressed explicitly and the conditional probability density $\mathcal{P}(x,t)$ can be solved exactly. For the general system, we discuss the modified force in detail and give a path integral representation in the appendix.

## 4. Results for two solvable models

In this section we consider the analytic results of the free particle system and the harmonic system. The modified forces in these two models had been presented in [32], but the explicit solutions of the conditional Langevin equation and the most probable path were not discussed there. Our purpose is to examine the consistence of the conditional Langevin equation with the conditional probability density, and to show the explicit expressions for the most probable paths in these two models.

### 4.1. Free particle system

For the free particle system $U(x)=0$, the propagator and the transition probability can be simply computed

$$\rho(x_e, t_{tr} | x_0, 0) = \langle x_e | e^{-t_{tr}\hat{H}} | x_0 \rangle = \sqrt{\frac{\beta m\gamma}{4\pi t_{tr}}} \exp\left\{-\beta \frac{1}{4} \frac{m\gamma}{t_{tr}} (x_e - x_0)^2\right\} \tag{24}$$

with the Hamiltonian $\hat{H} = -\frac{1}{\beta\gamma m} \frac{\partial^2}{\partial x^2}$. The conditional probability density $\mathcal{P}(x,t)$ is obtained straightforwardly

$$\mathcal{P}(x,t) = \sqrt{\frac{\beta m\gamma t_{tr}}{4\pi t(t_{tr}-t)}} \exp\left\{-\beta \frac{1}{4} m\gamma \frac{t_{tr}}{t(t_{tr}-t)} \left[x - \frac{x_0(t_{tr}-t) + x_e t}{t_{tr}}\right]^2\right\} . \tag{25}$$

Obviously $\mathcal{P}(x,t)$ is a Gaussian probability distribution, and the most probable position in the configurational space at time $t$ is the mean of $\mathcal{P}(x,t)$. This observation yields the expression for the most probable position as a function of $t$

$$x_m(t) = \frac{x_0(t_{tr}-t) + x_e t}{t_{tr}} . \tag{26}$$

As $t$ passes from 0 to $t_{tr}$, $x_m(t)$ goes from $x_0$ to $x_e$ with a constant velocity. We may demonstrate that $x_m(t)$ is actually the most probable path in the transition path space. One can easily examine that $x_m(t)$ in Eq. (26) is exactly the minimum action path in Eqs. (14)-(15) for the free particle system. Then it is clear that the minimum action principle for the transition probability in Eq. (11) works in the free particle system.

Let us turn to the conditional overdamped Langevin equation in Eq. (23). The modified force in Eq. (22) now becomes

$$-\frac{\partial \tilde{U}}{\partial x} = m\gamma \frac{x_e - x}{t_{tr} - t} . \tag{27}$$

The conditional overdamped Langevin equation is then

$$\frac{dx}{dt} = \frac{x_e - x}{t_{tr} - t} + \sqrt{\frac{2}{\beta m\gamma}} \eta(t) . \tag{28}$$

To exactly solve Eq. (28), We treat the white noise stochastic process $\eta(t)$ as a "function" of $t$. An explicit expression of the solution is easily given by

$$x(t) = \frac{x_0(t_{tr}-t) + x_e t}{t_{tr}} + \sqrt{\frac{2}{\beta m \gamma}} \int_0^t \frac{t_{tr}-t}{t_{tr}-s} \eta(s) ds \ . \tag{29}$$

The constraint $x(0) = x_0$, $x(t_{tr}) = x_e$ can be verified straightforwardly. Recall Eq. (2). The Wiener process guarantees that $\int_0^t \frac{t_{tr}-t}{t_{tr}-s} \eta(s) ds$ is a normal random variable, of which the mean is 0 and the variance is

$$\left\langle \left[ \int_0^t \frac{t_{tr}-t}{t_{tr}-s} \eta(s) ds \right]^2 \right\rangle = \int_0^t \int_0^t \frac{t_{tr}-t}{t_{tr}-s} \frac{t_{tr}-t}{t_{tr}-s'} \langle \eta(s) \eta(s') \rangle ds ds'$$

$$= \int_0^t \int_0^t \frac{t_{tr}-t}{t_{tr}-s} \frac{t_{tr}-t}{t_{tr}-s'} \delta(s-s') ds ds'$$

$$= \int_0^t \left( \frac{t_{tr}-t}{t_{tr}-s} \right)^2 ds \tag{30}$$

$$= \frac{t(t_{tr}-t)}{t_{tr}}.$$

Therefore, the probability distribution of the configurational space at time $t$ from Eq. (29), is the Gaussian distribution with the mean $\frac{x_0(t_{tr}-t) + x_e t}{t_{tr}}$ and the variance $\frac{2}{\beta m \gamma} \frac{t(t_{tr}-t)}{t_{tr}}$. This is exactly the conditional probability density $\mathcal{P}(x,t)$ in Eq. (25). Hence, we have approached to the statement that the conditional Langevin equation is consistent with the desired probability density, and is thus an effective tool to sample the transition path ensemble.

### 4.2. Harmonic system

The analysis of the harmonic system $U(x) = \frac{1}{2} m\omega^2 x^2$ is similar to that of the free particle system. First we look at the explicit expressions of the propagator and the transition probability

$$\left\langle x_{e}\left|e^{-t_{tr}\hat{H}}\right|x_{0}\right\rangle = \sqrt{\frac{\beta m\omega^{2}}{4\pi \sinh\left(\frac{\omega^{2}t_{tr}}{\gamma}\right)}}$$

$$\times \exp\left\{-\beta\frac{m\omega^{2}}{4\sinh\left(\frac{\omega^{2}t_{tr}}{\gamma}\right)}\left[\cosh\left(\frac{\omega^{2}t_{tr}}{\gamma}\right)\left(x_{0}^{2}+x_{e}^{2}\right)-2x_{0}x_{e}\right]+\frac{\omega^{2}t_{tr}}{2\gamma}\right\}, \quad (31)$$

$$\rho(x_{e},t_{tr}|x_{0},0) = \sqrt{\frac{\beta m\omega^{2}}{4\pi \sinh\left(\frac{\omega^{2}t_{tr}}{\gamma}\right)}}\exp\left\{-\beta\frac{m\omega^{2}}{4\sinh\left(\frac{\omega^{2}t_{tr}}{\gamma}\right)}\left[e^{-\frac{\omega^{2}t_{tr}}{2\gamma}}x_{0}-e^{\frac{\omega^{2}t_{tr}}{2\gamma}}x_{e}\right]^{2}+\frac{\omega^{2}t_{tr}}{2\gamma}\right\}$$

with the Hamiltonian $\hat{H} = -\frac{1}{\beta\gamma m}\frac{\partial^{2}}{\partial x^{2}}+\frac{\beta}{4\gamma m}\left[\left(m\omega^{2}x\right)^{2}-\frac{2}{\beta}m\omega^{2}\right]$. The conditional probability density $\mathcal{P}(x,t)$ is

$$\mathcal{P}(x,t) = \sqrt{\frac{\beta m\omega^{2}}{4\pi}\frac{\sinh\left(\frac{\omega^{2}t_{tr}}{\gamma}\right)}{\sinh\left(\frac{\omega^{2}t}{\gamma}\right)\sinh\left(\frac{\omega^{2}(t_{tr}-t)}{\gamma}\right)}}$$

$$\times\exp\left\{-\beta m\omega^{2}\frac{\sinh\left(\frac{\omega^{2}t_{tr}}{\gamma}\right)}{4\sinh\left(\frac{\omega^{2}t}{\gamma}\right)\sinh\left(\frac{\omega^{2}(t_{tr}-t)}{\gamma}\right)}\left[x-\frac{x_{0}\sinh\left(\frac{\omega^{2}(t_{tr}-t)}{\gamma}\right)+x_{e}\sinh\left(\frac{\omega^{2}t}{\gamma}\right)}{\sinh\left(\frac{\omega^{2}t_{tr}}{\gamma}\right)}\right]^{2}\right\}. \quad (32)$$

It is a Gaussian distribution and the most probable position is the mean

$$x_{m}(t) = \frac{x_{0}\sinh\left(\frac{\omega^{2}(t_{tr}-t)}{\gamma}\right)+x_{e}\sinh\left(\frac{\omega^{2}t}{\gamma}\right)}{\sinh\left(\frac{\omega^{2}t_{tr}}{\gamma}\right)}. \quad (33)$$

Eq. (33) is the solution of Eqs. (14)-(15) for the minimum action path. Now we see that the most probable path is also the minimum action path in the harmonic system.

For the conditional Langevin equation, the modified force is

$$-\frac{\partial \tilde{U}}{\partial x} = -\frac{m\omega^2}{\sinh\left(\frac{\omega^2(t_{tr}-t)}{\gamma}\right)}\left[\cosh\left(\frac{\omega^2(t_{tr}-t)}{\gamma}\right)x - x_e\right]. \tag{34}$$

The corresponding Langevin equation is then

$$\frac{dx}{dt} = -\frac{\omega^2}{\gamma \sinh\left(\frac{\omega^2(t_{tr}-t)}{\gamma}\right)}\left[\cosh\left(\frac{\omega^2(t_{tr}-t)}{\gamma}\right)x - x_e\right] + \sqrt{\frac{2}{\beta m \gamma}}\eta(t). \tag{35}$$

This equation had also been discussed in [54]. We can give the explicit solution of Eq. (35) like we do for the free particle system

$$x(t) = \frac{x_0 \sinh\left(\frac{\omega^2(t_{tr}-t)}{\gamma}\right) + x_e \sinh\left(\frac{\omega^2 t}{\gamma}\right)}{\sinh\left(\frac{\omega^2 t_{tr}}{\gamma}\right)} + \sqrt{\frac{2}{\beta m \gamma}}\int_0^t \frac{\sinh\left(\frac{\omega^2(t_{tr}-t)}{\gamma}\right)}{\sinh\left(\frac{\omega^2(t_{tr}-s)}{\gamma}\right)}\eta(s)ds. \tag{36}$$

The resulting probability distribution of $x(t)$ is a Gaussian distribution with the mean

$$\frac{x_0 \sinh\left(\frac{\omega^2(t_{tr}-t)}{\gamma}\right) + x_e \sinh\left(\frac{\omega^2 t}{\gamma}\right)}{\sinh\left(\frac{\omega^2 t_{tr}}{\gamma}\right)} \quad \text{and the variance}$$

$$\frac{2}{\beta m \gamma}\int_0^t \frac{\sinh^2\left(\frac{\omega^2(t_{tr}-t)}{\gamma}\right)}{\sinh^2\left(\frac{\omega^2(t_{tr}-s)}{\gamma}\right)}ds = \frac{2}{\beta m \omega^2}\sinh^2\left(\frac{\omega^2(t_{tr}-t)}{\gamma}\right)\coth\left(\frac{\omega^2(t_{tr}-s)}{\gamma}\right)\bigg|_{s=0}^{s=t}$$

$$= \frac{2}{\beta m \omega^2}\frac{\sinh\left(\frac{\omega^2(t_{tr}-t)}{\gamma}\right)\sinh\left(\frac{\omega^2 t}{\gamma}\right)}{\sinh\left(\frac{\omega^2 t_{tr}}{\gamma}\right)}. \tag{37}$$

Obviously, it is exactly $\mathcal{P}(x,t)$ in Eq. (32). Hence, the consistence of the conditional Langevin equation with the conditional probability density is proved.

Remark that the results of the harmonic system reduce to that of the free particle system in the limit $\omega \to 0$.

## 5. Conclusions

In this article we study the transition path ensemble of the system driven by overdamped Langevin equation. By using an imaginary time propagator, we establish a path integral formulation of the transition probability. The minimum action principle for the most probable path, and the conditional Langevin equation for the desired conditional probability density, are derived. Taking the free particle system and the harmonic system as two exactly solvable examples, we confirm that the conditional Langevin equation is exactly consistent with the conditional probability density in the transition, and the minimum action principle actually leads to the most probable path. Our work provides more theoretical investigations of the transition path ensemble.

For the transition path sampling in the general system, one may expect efficient algorithms for numerically solving the conditional Langevin equation. The key problem is the evaluation of the time-dependent modified force in Eq. (22). Since it is very difficult if not impossible to express the modified force explicitly, some approximations [31, 33, 35] have been proposed. Rather than the approximated methods, our interest is the application of path integral molecular dynamics to evaluate the modified force. In the appendix we give a path integral representation for the modified force in Eqs. (45)-(47). Obviously, employing path integral molecular dynamics techniques to calculate $\left\langle \frac{\partial V}{\partial x_i} \right\rangle_P$ requires to evaluate the average in every time step of the trajectory $(x,t)$, thus is very time consuming. One possible approach is the adiabatic path integral molecular dynamics method [55-57]. Notice that $x$ and $t$ in the trajectory change slowly. It is natural to separate the time scale of $x$ from that of the other path integral beads $(x_2, \cdots, x_P)$

[or $(y_2,\cdots,y_P)$] in Eq. (44), e.g., by redefining much more smaller effective masses for $(x_2,\cdots,x_P)$. We expect $\left\langle \frac{\partial V}{\partial x_i} \right\rangle_P$ can be calculated more efficiently in the adiabatic representation. This adiabatic path integral molecular dynamics technique deserves further test.

**Appendix. Path integral representation of the modified force**

The modified force of the conditional Langevin equation can be explicitly expressed in some solvable cases. In Section 4 we show the results of the free particle system and the harmonic system. Here we discuss the modified force of the general system.

The modified force is defined in Eq. (22) with the Hamiltonian in Eq. (6). Notice that we can express the modified force as

$$-\frac{\partial \tilde{U}}{\partial x} = \frac{2}{\beta} \frac{\frac{\partial}{\partial x}\left\langle x_e \left| e^{-(t_{tr}-t)\hat{H}} \right| x \right\rangle}{\left\langle x_e \left| e^{-(t_{tr}-t)\hat{H}} \right| x \right\rangle}. \tag{38}$$

We may employ the popular splitting method called Trotter splitting [58] to the evolution operator $e^{-(t_{tr}-t)\hat{H}}$, which yields

$$e^{-\Delta t \hat{H}} \approx e^{-\frac{\Delta t}{2}\hat{V}} e^{-\Delta t \hat{T}} e^{-\frac{\Delta t}{2}\hat{V}} \tag{39}$$

for a small $\Delta t$. Here $\hat{T}$ and $\hat{V}$ are defined in Eq. (7). Splitting $t_{tr}-t$ into $P$ intervals and taking the limit $P \to \infty$ leads to the expression

$$\left\langle x_e \left| e^{-(t_{tr}-t)\hat{H}} \right| x \right\rangle = \lim_{P\to\infty} \left\langle x_e \left| \left( e^{-\frac{t_{tr}-t}{2P}\hat{V}} e^{-\frac{t_{tr}-t}{P}\hat{T}} e^{-\frac{t_{tr}-t}{2P}\hat{V}} \right)^P \right| x \right\rangle$$

$$= \lim_{P\to\infty} \frac{1}{\mathcal{N}} \int dx_2 \cdots dx_P \exp\left\{ -\frac{\beta m \gamma P}{4(t_{tr}-t)}\left[(x_e-x_2)^2 + \cdots + (x_P-x)^2\right] \right. \tag{40}$$

$$\left. -\frac{t_{tr}-t}{P}\left[ V(x_2) + \cdots + V(x_P) + \frac{1}{2}V(x_e) + \frac{1}{2}V(x) \right] \right\}.$$

Here $(x_e, x_2, \cdots, x_P, x)$ is the discrete time trajectory and $\mathcal{N}$ is a constant independent of $x$.

Now we adopt the open path staging coordinate transformation [59]

$$y_1 = \frac{1}{2}(x_e + x),$$
$$y_s = x_s - \frac{1}{s}\left[(s-1)x_{s+1} + x_1\right], \quad s = 2, \cdots, P \tag{41}$$
$$y_{P+1} = x_e - x$$

to simplify the derivation, where we set $x_1 = x_e$ and $x_{P+1} = x$. The inverse transformation is

$$x_1 = y_1 + \frac{1}{2} y_{P+1},$$
$$x_s = y_1 + \frac{P/2 - s + 1}{P} y_{P+1} + \sum_{t=s}^{P} \frac{s-1}{t-1} y_t, \quad s = 2, \cdots, P \tag{42}$$
$$x_{P+1} = y_1 - \frac{1}{2} y_{P+1}.$$

Eq. (40) becomes

$$\langle x_e | e^{-(t_{tr}-t)\hat{H}} | x \rangle = \lim_{P \to \infty} \frac{1}{\mathcal{N}} \int dy_2 \cdots dy_P \exp\left\{-\frac{\beta m \gamma P}{4(t_{tr}-t)} \left[\sum_{i=2}^{P} \frac{i}{i-1} y_i^2 + \frac{1}{P} y_{P+1}^2\right]\right.$$
$$\left. - \frac{t_{tr}-t}{P}\left[V(x_2(y)) + \cdots + V(x_P(y)) + \frac{1}{2}V(x_e) + \frac{1}{2}V(x)\right]\right\}. \tag{43}$$

The integral variables transform from $(x_2, \cdots, x_P)$ to $(y_2, \cdots, y_P)$. Then we have

$$\frac{\partial}{\partial x} \langle x_e | e^{-(t_{tr}-t)\hat{H}} | x \rangle$$
$$= \frac{1}{2}\frac{\partial}{\partial y_1} \langle x_e | e^{-(t_{tr}-t)\hat{H}} | x \rangle - \frac{\partial}{\partial y_{P+1}} \langle x_e | e^{-(t_{tr}-t)\hat{H}} | x \rangle$$
$$= \lim_{P \to \infty} \frac{1}{\mathcal{N}} \int dy_2 \cdots dy_P \exp\left\{-\frac{\beta m \gamma P}{4(t_{tr}-t)}\left[\sum_{i=2}^{P} \frac{i}{i-1} y_i^2 + \frac{1}{P} y_{P+1}^2\right] - \frac{t_{tr}-t}{P}\left[V(x_2(y)) + \cdots + V(x_P(y)) + \frac{1}{2}V(x_e) + \frac{1}{2}V(x)\right]\right\}$$
$$\times \left\{\frac{\beta m \gamma}{2} \frac{x_e - x}{t_{tr}-t} - \frac{t_{tr}-t}{P}\left[\frac{1}{2}\frac{\partial V}{\partial x} + \sum_{i=2}^{P} \frac{i-1}{P} \frac{\partial V}{\partial x_i(y)}\right]\right\}$$
$$= \lim_{P \to \infty} \frac{1}{\mathcal{N}} \int dx_2 \cdots dx_P \exp\left\{-\frac{\beta m \gamma P}{4(t_{tr}-t)}\left[(x_e - x_2)^2 + \cdots + (x_P - x)^2\right] - \frac{t_{tr}-t}{P}\left[V(x_2) + \cdots + V(x_P) + \frac{1}{2}V(x_e) + \frac{1}{2}V(x)\right]\right\}$$
$$\times \left\{\frac{\beta m \gamma}{2} \frac{x_e - x}{t_{tr}-t} - \frac{t_{tr}-t}{P}\left[\frac{1}{2}\frac{\partial V}{\partial x} + \sum_{i=2}^{P} \frac{i-1}{P} \frac{\partial V}{\partial x_i}\right]\right\}.$$
$$\tag{44}$$

In the last step of Eq. (44) we perform the inverse coordinate transformation. It is straightforward to show

$$\frac{\frac{\partial}{\partial x}\left\langle x_e \left| e^{-(t_{tr}-t)\hat{H}} \right| x \right\rangle}{\left\langle x_e \left| e^{-(t_{tr}-t)\hat{H}} \right| x \right\rangle} = \frac{\beta m \gamma}{2} \frac{x_e - x}{t_{tr} - t} - \lim_{P \to \infty} \frac{t_{tr} - t}{P} \frac{1}{2} \frac{\partial V}{\partial x} - \lim_{P \to \infty} \sum_{i=2}^{P} \frac{t_{tr} - t}{P} \frac{i-1}{P} \left\langle \frac{\partial V}{\partial x_i} \right\rangle_P$$

$$= \frac{\beta m \gamma}{2} \frac{x_e - x}{t_{tr} - t} - \lim_{P \to \infty} \sum_{i=2}^{P} \frac{t_{tr} - t}{P} \frac{i-1}{P} \left\langle \frac{\partial V}{\partial x_i} \right\rangle_P \quad (45)$$

where $\langle \ \rangle_P$ denotes the average over the $P$-splitting path integral representation. Notice that for trajectory $(x_e, x_2, \cdots, x_P, x)$ the discrete time is $t_i = t_{tr} - \frac{i-1}{P}(t_{tr} - t)$, and we can express the second term in the right-hand side of Eq. (45) as

$$\lim_{P \to \infty} \sum_{i=2}^{P} \frac{t_{tr} - t}{P} \frac{i-1}{P} \left\langle \frac{\partial V}{\partial x_i} \right\rangle_P = \lim_{P \to \infty} \sum_{i=2}^{P} \frac{t_{tr} - t}{P} \frac{t_{tr} - t_i}{t_{tr} - t} \left\langle \frac{\partial V}{\partial x_i} \right\rangle_P = \int_t^{t_{tr}} \frac{t_{tr} - s}{t_{tr} - t} \left\langle \frac{\partial V}{\partial x(s)} \right\rangle ds \ . \quad (46)$$

Here $\langle \ \rangle$ is the average over the path space from $x$ at $t$ to $x_e$ at $t_{tr}$. Then the modified force can be presented from Eqs. (38), (45) and (46)

$$-\frac{\partial \tilde{U}}{\partial x} = m \gamma \frac{x_e - x}{t_{tr} - t} - \frac{2}{\beta} \int_t^{t_{tr}} \frac{t_{tr} - s}{t_{tr} - t} \left\langle \frac{\partial V}{\partial x(s)} \right\rangle ds \ . \quad (47)$$

This formulation had been proposed in [35] using a different approach.

## Statements and Declarations